# Quasi-Vertical β-$Ga_2O_3$ Schottky Diodes on Sapphire Using All-LPCVD Growth and Plasma-Free Ga-Assisted Etching

Saleh Ahmed Khan[1], Ahmed Ibreljic[1], A F M Anhar Uddin Bhuiyan[1, a)]

[1]Department of Electrical and Computer Engineering, University of Massachusetts Lowell, MA 01854, USA

[a)] Corresponding author Email: anhar_bhuiyan@uml.edu

**Abstract**

This work demonstrates quasi-vertical β-$Ga_2O_3$ Schottky barrier diodes (SBDs) fabricated on c-plane sapphire substrates using an all-low-pressure chemical vapor deposition (LPCVD)-based, plasma-free process flow that integrates both epitaxial growth of high-quality β-$Ga_2O_3$ heteroepitaxial film with an *in situ* Ga-assisted β-$Ga_2O_3$ etching. A 6.3 µm thick ($\bar{2}01$) oriented β-$Ga_2O_3$ epitaxial layer structure was grown on c-plane sapphire with 6° miscut, comprising a moderately Si-doped (2.1×$10^{17}$ $cm^{-3}$) 3.15 µm thick drift layer and a heavily doped (1×$10^{19}$ $cm^{-3}$) contact layer on an unintentionally doped buffer layer. Mesa isolation was achieved via Ga-assisted plasma-free LPCVD etching, producing a ~60° inclined mesa sidewalls with an etch depth of 3.6 µm. The fabricated SBDs exhibited excellent forward current-voltage characteristics, including a turn-on voltage of 1.22 V, an ideality factor of 1.29, and a Schottky barrier height of 0.83 eV. The minimum differential specific on-resistance was measured to be 8.6 mΩ·$cm^2$, and the devices demonstrated high current density capability (252 A/$cm^2$ at 5 V). Capacitance-voltage analysis revealed a net carrier concentration of 2.1 × $10^{17}$ $cm^{-3}$, uniformly distributed across the β-$Ga_2O_3$ drift layer. Temperature-dependent J-V-T measurements from 25 °C to 250 °C revealed thermionic emission-dominated transport with strong thermal stability. The Schottky barrier height increased from 0.80 eV to 1.16 eV, and the ideality factor rose modestly from 1.31 to 1.42 over this temperature range. Reverse leakage current remained low, increasing from ~5 × $10^{-6}$ A/$cm^2$ at 25 °C to ~1 × $10^{-4}$ A/$cm^2$ at 250 °C, with the $I_{on}/I_{off}$ ratio decreasing from ~1 × $10^7$ to 5 × $10^5$. The devices achieved breakdown voltages ranging from 73 V to 100 V, corresponding to parallel-plate electric field strengths of 1.66 - 1.94 MV/cm. These results highlight the potential of LPCVD-grown and etched β-$Ga_2O_3$ devices for high-performance, thermally resilient power electronics applications.

## I. Introduction

β-$Ga_2O_3$ stands out as a promising ultra-wide bandgap semiconductor for next-generation high-voltage, high-efficiency power electronics, due to its ultrawide bandgap (~4.8 eV), high breakdown electric field (6-8 MV/cm), and the availability of low-cost, large-area native substrates grown by melt-based growth methods [1]. These characteristics result in a Baliga's figure of merit that surpasses those of conventional wide bandgap power semiconductors such as SiC and GaN, making β-$Ga_2O_3$ an ideal candidate for next-generation power conversion systems, radio-frequency amplification, and extreme-environment electronics [2-7]. Substantial progress in high-quality β-$Ga_2O_3$ epitaxial film growth has enabled the realization of both lateral and vertical device architectures with excellent breakdown strength and thermal stability [8-35]. To fully leverage the potential of β-$Ga_2O_3$ in high-power vertical device architectures, precise control over vertical isolation and trench formation is essential. This requires fabrication techniques capable of defining deep, high-aspect-ratio structures with minimal damage and high fidelity. However, the absence of p-type doping in β-$Ga_2O_3$ necessitates electric field control through mesa, trench, or fin geometries, structures that are highly sensitive to processing-induced surface degradation. Conventional plasma-based dry etching techniques can produce anisotropic profiles in β-$Ga_2O_3$ but often lead to lattice distortion and damage at the sidewalls of etched regions [36, 37], which can degrade overall device performance. To reduce such damage, alternative approaches including wet chemical etching [38-40], metal-assisted chemical etching (MacEtch) [41, 42], and *in situ* etching in MBE [43, 44], MOCVD [45], and HVPE [46-48] systems have been explored. We recently developed an innovative Ga-assisted LPCVD based *in situ* etching technique using solid-source metallic Ga as etchant, which enables anisotropic and plasma-free patterning of β-$Ga_2O_3$ [49]. This technique exploits a thermally activated surface reaction in an oxygen-deficient low-pressure CVD environment, leading to the formation of volatile $Ga_2O$ suboxides that facilitate selective material removal. The etch chemistry: $4Ga + Ga_2O_3 \rightarrow 3Ga_2O$ proceeds cleanly under LPCVD conditions, avoiding the formation of plasma-induced surface states or subsurface defects. Through systematic studies, we demonstrated that the etch rate is tunable by controlling the Ga source-to-substrate distance, process temperature, and carrier gas flow rate, with peak etch rates exceeding ~2.25 μm/hr under optimized conditions. In addition to achieving high etch rates, our LPCVD-based etching technique also exhibited a strong crystallographic selectivity and in-plane anisotropy. On (010)-oriented β-$Ga_2O_3$ substrates, trenches aligned along the (100) orientation yielded the most stable sidewalls, smooth, vertical, and with minimal lateral undercut, indicating that LPCVD based metallic Ga-assisted etching can achieve orientation-controlled 3D structures, essential for high power β-$Ga_2O_3$ vertical devices.

Building upon this foundational etching technology in LPCVD environment, this work demonstrates the LPCVD-grown ($\bar{2}01$) β-$Ga_2O_3$ quasi-vertical Schottky barrier diodes (SBDs) fabricated on insulating c-plane sapphire substrates, utilizing a fully plasma-free LPCVD growth and etching process. While β-$Ga_2O_3$ native substrates offer excellent lattice matching for developing high-quality epitaxial drift layer, their widespread adoption is still limited by high substrate cost and inherently low thermal conductivity. Sapphire, in contrast, combines excellent electrical isolation from its ultrawide bandgap, mechanical robustness, relatively higher thermal conductivity, and significantly lower cost, making it an attractive platform for scalable power devices. One potential drawback of using sapphire could be its lattice mismatch with β-$Ga_2O_3$. However, this challenge is also addressed by orienting β-$Ga_2O_3$ growth along its ($\bar{2}01$) orientation, which supports high-quality epitaxial film formation. Additionally, the use of offcut c-plane sapphire substrates (6° miscut) further promotes step-flow growth with smoother surface morphology. Our LPCVD technique has already demonstrated excellent film quality with high growth rates, as evidenced by our recent work where electron mobilities up to 149 $cm^2/V\cdot s$ at carrier concentration of $1.15\times10^{17}$ $cm^{-3}$ were achieved in β-$Ga_2O_3$ films grown on off-axis sapphire substrates [10]. In this work, we integrate the high-quality LPCVD-grown β-$Ga_2O_3$ epitaxy with *in situ* Ga-assisted LPCVD etching process to realize quasi-vertical Schottky barrier diodes on insulating c-plane sapphire substrates- all within a fully plasma-free process flow. This combined approach not only minimizes interface damage but also simplifies process integration and reduces fabrication overhead.

## II. Experimental Details

### II. A. LPCVD Growth of β-$Ga_2O_3$ Epitaxial Stack:

The ($\bar{2}01$) β-$Ga_2O_3$ epitaxial layers were grown in a custom-designed LPCVD chamber on c-plane sapphire substrates with a 6° miscut. This intentional off-axis orientation facilitates step-flow growth and suppresses defect formation, enabling high crystalline quality of the epitaxial film [10, 28, 50]. Prior to growth, the substrate was sequentially cleaned with acetone, isopropyl alcohol (IPA), and deionized (DI) water, followed by nitrogen blow-drying. Ultra-high purity argon (99.9999%) was used as both the carrier and purge gas, while oxygen (99.999%) and metallic gallium pellets (99.99999%) served as the oxygen and gallium sources, respectively. Silicon tetrachloride ($SiCl_4$) was introduced as the n-type Si dopant, with flow rates adjusted to achieve desired doping concentrations. The growth was carried out at a substrate temperature of 1000 °C and a chamber pressure of ~1.5 Torr, with the substrate placed 7 cm from the gallium source. The epitaxial stack consisted of three distinct layers with a total thickness of 6.3 µm. Growth began with a 10-minute unintentionally doped (UID) $\beta$-$Ga_2O_3$ buffer layer (1.05 µm thick) grown without any $SiCl_4$ flow. This was followed by a 20-minute growth of $n^+$ Si-doped $\beta$-$Ga_2O_3$ layer (2.10 µm thick) using a $SiCl_4$ flow rate of 0.5 sccm to achieve a high doping concentration of $1\times10^{19}$ cm$^{-3}$, forming the contact layer for the Schottky barrier diode. Finally, the Si doped $\beta$-$Ga_2O_3$ drift layer was grown for 30 minutes using a reduced $SiCl_4$ flow rate of 0.001 sccm, targeting a doping concentration of $2.1 \times 10^{17}$ cm$^{-3}$ and achieving a total thickness of 3.15 µm.

**II. B. Plasma-Damage Free Etching of $\beta$-$Ga_2O_3$ using LPCVD:**

Plasma-free mesa isolation was achieved using our custom-built horizontal LPCVD system, where etching is driven by the thermal reaction between $\beta$-$Ga_2O_3$ and upstream metallic Ga vapor in an oxygen-deficient environment. The etch process was conducted at 1050 °C and a pressure of ~1.2 Torr, with the sample placed 2 cm downstream from the solid Ga source. Ultra-high purity argon (99.9999%) was used as the carrier and purge gas. Prior to etching, the samples were cleaned using acetone, IPA, and DI water, followed by nitrogen blow-drying. A 100 nm-thick $SiO_2$ hard mask was deposited using plasma-enhanced chemical vapor deposition (PECVD) and patterned via optical lithography to expose the regions targeted for etching. The samples were etched for 2 hours and 15 minutes, resulting in a measured etch depth of 3.6 µm, corresponding to an average etch rate of 1.6 µm/hr. This etch depth was intentionally selected to fully remove the drift layer and selectively reach into the $n^+$ region, enabling device isolation and providing access to the $n^+$ layer for subsequent ohmic contact formation. After etching, the $SiO_2$ mask was removed using a 1:50 diluted buffered oxide etch (BOE).

### II. C. Structural, Surface Morphological, and Chemical Characterization

The surface morphology of the β-$Ga_2O_3$ epi-stack was characterized using a JEOL JSM-7401F field-emission scanning electron microscope (FESEM) to visualize surface texture and etch profiles, and a Park Systems XE-100 atomic force microscope (AFM) operated in tapping mode to quantify nanoscale roughness. The crystalline structure, phase purity, and orientation of the films were evaluated by high-resolution X-ray diffraction (XRD) using a Rigaku SmartLab diffractometer equipped with a Cu Kα radiation source ($\lambda$ = 1.5418 Å). Wide-angle 2θ–ω scans and rocking curves (ω-scans) were used to confirm phase purity and determine the out-of-plane crystalline orientation. Raman spectroscopy was carried out using a Horiba LabRam Evolution Raman spectrometer with a 532 nm excitation source to assess phonon modes and verify structural integrity before and after etching. X-ray photoelectron spectroscopy (XPS) measurements were performed using a Thermo Scientific XPS system with a monochromated Al Kα X-ray source (E = 1486.6 eV) to analyze chemical composition and bonding states. High-resolution scans of the Ga 3s and O 1s core levels were used to calculate the O/Ga ratio and monitor potential changes in surface chemistry following the LPCVD-based etching process.

### II. D. Quasi-Vertical β-$Ga_2O_3$ Schottky Diode Fabrication

Following mesa isolation by LPCVD Ga-assisted etching, device fabrication was completed using sequential photolithography, passivation, metallization, and annealing steps. First, a 200 nm-thick $SiO_2$ passivation layer was deposited across the entire wafer surface using PECVD. Photolithography and buffered oxide etch (BOE) were then used to open windows in the $SiO_2$ layer to define the cathode region. A Ti/Au (20 nm/50 nm) metal stack was deposited by electron-beam evaporation and patterned via lift-off to form the cathode contact on the β-$Ga_2O_3$ surface surrounding the recessed mesa. This was followed by rapid thermal annealing (RTA) at 470 °C for 1 min in nitrogen ambient to improve metal-semiconductor contact characteristics. Subsequently, a second photolithography step was used to define the anode contact region. Ni/Au (30 nm/50 nm) was deposited via electron-beam evaporation and patterned using lift-off to form the Schottky contact on top of the etched mesa. The final device structure and full fabrication process flow are illustrated in Figs. 1(a) and (d), with the FESEM and AFM images (Figs. 1(b) and (c)) confirming an etch depth of 3.6 μm with inclined (~60°) sidewalls. Post-fabrication electrical characterization, including room-temperature and high-temperature (up to 250 °C) Current-Voltage (J-V), Capacitance-Voltage (C-V) and reverse breakdown measurements, was performed using a Keithley 4200A-SCS Semiconductor Parameter Analyzer.

## III. Results and Discussions

To evaluate the material quality of the heteroepitaxial β-$Ga_2O_3$ epitaxial stack used for device fabrication, comprehensive structural and morphological characterization was performed. Figure 2(a) shows a top-view FESEM image that reveals distinct step-flow surface morphology, indicative of epitaxial layer-by-layer growth. The surface AFM scan in Fig. 2(b) shows uniform morphology with step-flow features and an RMS roughness of 4.7 nm, confirming a smooth and well-ordered surface. These morphological features are characteristic of step-flow growth on off-axis sapphire substrates, consistent with our prior work on LPCVD-grown high quality ($\bar{2}01$) β-$Ga_2O_3$ films with similar doping levels [10].

High-resolution X-ray diffraction (XRD) analysis further confirmed the film's structural quality and phase purity. As shown in Fig. 2(c), strong diffraction peaks corresponding exclusively to the ($\bar{h}0l$) family of monoclinic β-$Ga_2O_3$ planes are observed, with no evidence of secondary phases. The presence of higher-order ($\bar{2}01$) peaks such as ($\bar{4}02$), ($\bar{6}03$), and ($\bar{8}04$) signifies coherent long-range ordering and crystallographic orientation fidelity. The rocking curve (ω-scan) of the ($\bar{4}02$) reflection, shown in Fig. 2(d), exhibits a narrow full width at half maximum (FWHM) of 287 arcsec. This value is among the lowest reported for heteroepitaxial β-$Ga_2O_3$ film on sapphire, typically achieved using epitaxial techniques such as MOCVD and MBE [50-54].

To evaluate the structural integrity of the heteroepitaxial β-$Ga_2O_3$ film before and after Ga-assisted *in situ* LPCVD etching, Raman spectroscopy was performed on both the pristine (unetched) and etched surfaces, as shown in Fig. 3. The Raman spectra exhibit characteristic vibrational modes corresponding to monoclinic ($\bar{2}01$) β-$Ga_2O_3$, with distinct peaks attributed to both $A_g$ and $B_g$ phonon symmetries. These modes arise from the 30 phonon modes predicted by group theory for monoclinic β-$Ga_2O_3$, of which 27 are optical and 3 are acoustic. The optical modes are classified as 10 $A_g$, 5 $B_g$, 4 $A_u$, and 8 $B_u$, where the $A_g$ and $B_g$ modes are Raman-active, and the $A_u$ and $B_u$ modes are infrared-active [55]. In the spectra of the heteroepitaxial ($\bar{2}01$) β-$Ga_2O_3$ films, distinct peaks are observed at 114.1, 144.3, 474.1, and 652.5 $cm^{-1}$, corresponding to $B_g$ modes, and at 169.3, 200.1, 319.3, 347.1, 416.2, 630.5, and 766.7 $cm^{-1}$, corresponding to $A_g$ modes. These peak positions are in good agreement with previous experimental and theoretical studies [50, 55]. Importantly, the Raman signatures from the etched surface show no shift in peak position, no broadening, and no suppression of the fundamental vibrational modes compared to the pristine surface. This indicates that the LPCVD-based Ga-assisted etching process preserves the lattice symmetry and does not introduce structural degradation, rotational domains or phase transformation. The intensity of sapphire-related modes remains similar between the two spectra, further confirming that the optical penetration depth remains consistent and the observed phonon modes are intrinsic to the β-$Ga_2O_3$ layer. The retention of both $A_g$ and $B_g$ modes with sharp and symmetric profiles indicates good crystalline quality of both pristine and etched film surface.

To investigate the surface chemistry before and after etching, X-ray photoelectron spectroscopy (XPS) was performed on both the pristine and etched β-$Ga_2O_3$ surfaces, as shown in Fig. 4. The wide-range survey spectra (Fig. 4(a)) reveal distinct peaks corresponding to Ga and O, with no detectable metallic contamination, confirming the chemical purity of both surfaces. Notably, no silicon-related peaks were detected on the etched surface within the sensitivity limits of XPS, suggesting that any potential Si diffusion from the $SiO_2$ mask into the β-$Ga_2O_3$ surface might not occur under etching process conditions. High-resolution scans of the Ga 3s and O 1s regions were used for quantitative analysis, as shown in Figs. 4 (b-e). High-resolution O 1s spectra were deconvoluted into two components: a primary peak centered near 530.9 eV corresponding to lattice oxygen (Ga–O), and a secondary peak near 532.3 eV attributed to surface hydroxyl groups (O–H). The Ga 3s peak remained sharp and symmetric, indicating stable Ga–O bonding. Using sensitivity-factor-corrected areas of the O 1s and Ga 3s peaks, the O/Ga atomic ratio was calculated to be 1.49 (Oxygen: 59.81%, Gallium: 40.19%) for the unetched surface and 1.45 (Oxygen: 59.25%, Gallium: 40.75%) for the etched surface. These results indicate that both the pristine and etched surfaces retain their chemical integrity and near-stoichiometric composition.

Following the comprehensive structural, morphological, and chemical characterization of the LPCVD-grown β-$Ga_2O_3$ epitaxial stack, the electrical performance of the fabricated Schottky barrier diodes was evaluated to further assess the impact of the LPCVD-grown and etched β-$Ga_2O_3$ layers on device behavior. The current-voltage (J-V) characteristics of the diode, shown in Fig. 5(a), demonstrate excellent rectifying behavior with a clear exponential increase in forward current and a low leakage current in reverse bias. The device exhibits a low turn-on voltage ($V_{turn\text{-}on}$) of 1.22 V (assuming an on-state current density of 1 A $cm^{-2}$) and an ideality factor ($\eta$) of 1.29, indicating near-ideal thermionic emission transport across the metal-semiconductor junction. The extracted Schottky barrier height ($\Phi_B$) of 0.83 eV is consistent with typical Ni/$\beta$-$Ga_2O_3$ contacts and reflects good interface quality. The minimum differential specific on-resistance ($R_{on,sp}$) was determined to be 8.6 m$\Omega$·$cm^2$. This low $R_{on,sp}$ value confirms efficient current transport through the epitaxial stack and effective contact formation on the recessed $n^+$ $\beta$-$Ga_2O_3$ layer. The high current density capability of 252 A/$cm^2$ at 5 V further highlights the advantages of the excellent material quality achieved through LPCVD-grown $\beta$-$Ga_2O_3$. It is important to note that etch-induced damage can significantly degrade diode performance, affecting leakage current, ideality factor, and on-resistance. Conventional plasma-based reactive ion etching (RIE) techniques have been reported to introduce structural and chemical damage in $\beta$-$Ga_2O_3$, often necessitating additional post-etch treatments to restore interface quality [56 57]. In contrast, the Ga-assisted *in situ* LPCVD etching approach employed in this work offers a damage-free, thermally driven alternative that avoids ion bombardment and is expected to preserve the integrity of the etched sidewalls. Such integration of LPCVD growth and plasma-free etching within a single process flow is reflected in the low ideality factor and good rectification behavior observed in our devices.

To further evaluate the doping and junction properties, C-V measurements were performed, as shown in Fig. 5(b). The capacitance decreased with increasing reverse bias, and the corresponding $1/C^2$–V plot exhibited a linear relationship, characteristic of a uniformly doped Schottky junction. The $V_{bi}$ and $N_d^+$ - $N_a^-$ are determined using the following formulas using the relative permittivity of $\beta$-$Ga_2O_3$, $\varepsilon_r$ = 10, density of states in the conduction band, $N_C$ = $5.2\times10^{18}$ $cm^{-3}$, where A representing the device area [58, 59].

$$N_d^+ - N_a^- = \frac{2}{q\varepsilon_r\varepsilon_0 A^2\left(\frac{d\frac{1}{C^2}}{dV}\right)} \qquad (1)$$

$$\frac{A^2}{C^2} = qV_{bi} + \frac{kT}{q} In[\frac{N_c}{N_d^+ - N_a^-}] \qquad (2)$$

The built-in potential ($V_{bi}$), estimated by extrapolating the linear region of the $1/C^2$–V curve to the voltage axis, was found to be ~3.67 V. From the C-V analysis, an average net carrier density ($N_d^+ - N_a^-$) of $2.1 \times 10^{17}$ cm$^{-3}$ was extracted, matching well with the doping level targeted for the ($\bar{2}01$) β-$Ga_2O_3$ drift layer. As shown in Figure 5(c), the net carrier density profile is flat and uniform across the drift layer, confirming the consistency of doping achieved during the epitaxial growth process. Additionally, no clear indication of an upward slope or surface donor accumulation is observed in the carrier density profile, suggesting that the high-temperature *in situ* LPCVD etching process did not lead to any noticeable dopant diffusion from the $SiO_2$ mask, consistent with prior studies reporting minimal interdiffusion at β-$Ga_2O_3$/$SiO_2$ interfaces even after annealing at high temperatures [60].

To assess the thermal stability and transport mechanisms of the β-$Ga_2O_3$ Schottky barrier diodes fabricated on sapphire substrates, temperature-dependent current–voltage (J-V-T) measurements were performed from 25 °C to 250 °C, as shown in Fig. 6. In the linear-scale forward J-V plots shown in Fig. 6(a), the forward current density increases with temperature, which is characteristic of thermionic emission over the Schottky barrier. As the temperature rises, electrons gain additional thermal energy, increasing their probability of surmounting the Schottky barrier and resulting in enhanced forward conduction. This temperature-enhanced barrier injection becomes particularly prominent at low-to-moderate forward bias, where transport is barrier-limited and governed by the Schottky diode equations:

$$J = J_s\,[\exp(\frac{qV}{\eta k_0 T}) - 1] \qquad (3)$$

$$J_s = A^* T^2 \exp(-\frac{q\Phi_B}{k_0 T}) \qquad (4)$$

$$A^* = \frac{4\pi q m_n^* k_0^2}{h^3} \quad (5)$$

where q is the electric charge, $k_0$ is the Boltzmann constant, and η is the ideality factor, $J_s$ is the reverse saturation current density, $\Phi_B$ is the Schottky barrier height, and A* is Richardson's constant, which is calculated to be 41.04 A $cm^{-2}$ $K^{-2}$ [59, 61, 62]. The semi-logarithmic forward J–V characteristics (Fig. 6b) further illustrate this trend, revealing an increase in on-current from 53 to 90 A $cm^{-2}$ from 25 °C to 250 °C at the same forward bias of 4V. Notably, while the current continues to rise with temperature in the barrier-limited regime (low bias), the increase saturates in the high bias (ohmic) region. This saturation is likely due to increased phonon scattering and reduced carrier mobility within the drift region and contact layers at elevated temperatures, a behavior typical in semiconductors where the electron-phonon interaction dominates at high fields and temperatures [63]. Additionally, self-heating and series resistance effects may begin to influence the I-V shape at high current levels [61], particularly in devices with small footprint and limited thermal sinking. Figure 6(c) quantifies the temperature dependence of $R_{on,sp}$ extracted from the linear region of the forward J-V characteristics. A non-monotonic trend is observed: while $R_{on,sp}$ initially decreases with increasing temperature in the low-to-moderate forward bias regime due to thermally assisted carrier injection, the inset reveals that $R_{on,sp}$ increases in the high-bias regime at elevated temperatures. This divergent behavior arises from two competing effects. At low biases, thermionic emission dominates and benefits from enhanced carrier activation and interface injection, which reduces the effective series resistance. However, at higher forward biases where the current becomes limited by the series resistance of the drift region, the dominant factor becomes carrier mobility. As the temperature increases, enhanced lattice vibrations increase electron–phonon scattering, leading to reduced electron mobility in the β-$Ga_2O_3$ drift region and thus a rise in $R_{on,sp}$ under high-field conditions. This trend, decreasing $R_{on,sp}$ at low bias and increasing $R_{on,sp}$ at high bias with temperature, is characteristic of β-$Ga_2O_3$ Schottky barrier diodes and is consistent with earlier observations in radiation and thermally stressed devices[24, 59, 64]. The ability to maintain rectifying behavior and consistent conduction characteristics over a wide temperature range, demonstrates the structural and electrical robustness of the LPCVD-grown β-$Ga_2O_3$ layers and the reliability of the plasma-free device processing strategy.

Figure 7 summarizes the extracted Schottky barrier height ($\Phi_B$), ideality factor (η), reverse leakage current ($I_{reverse}$), and rectification ratio ($I_{on}/I_{off}$) as functions of temperature for the fabricated diodes. As shown in Fig. 7(a), the barrier height increases from 0.80 eV at 25 °C to 1.16 eV at 250 °C, consistent with the expectations of the thermionic emission (TE) model in the presence of barrier inhomogeneities [65-69]. At lower temperatures, current conduction tends to be dominated by electrons traversing lower-barrier patches at the metal/semiconductor interface. As temperature increases, more carriers acquire sufficient thermal energy to surmount higher barrier regions, effectively raising the extracted $\Phi_B$. This phenomenon, attributed to lateral inhomogeneities in the Schottky contact, has been reported in previous literatures for ultra-wide bandgap semiconductors [59, 70, 71], and is typically associated with non-idealities, including interfacial disorder, grain boundaries, or residual contaminants. The extracted ideality factor η, also shown in Fig. 7(a), increases modestly from 1.31 at 25 °C to 1.42 at 250 °C. This slight increase with temperature suggests a gradual deviation from ideal thermionic emission behavior, potentially due to enhanced recombination or tunneling contributions at higher temperatures. In practice, temperature-induced changes in η can also reflect evolving interface conditions, including thermally activated trap-assisted transport or bias-dependent modulation of interface states. Nevertheless, the relatively low η across the full temperature range indicates a high-quality Schottky junction with minimal leakage paths and well-behaved transport characteristics. Figure 7(b) further evaluates the thermal stability of the diode by plotting the reverse leakage current (measured at -4 V) and the $I_{on}/I_{off}$ ratio (measured at ±4 V) as functions of temperature. As expected, reverse leakage current increases with temperature, rising from $4.9 \times 10^{-6}$ A/cm² at 25 °C to $1.7 \times 10^{-4}$ A/cm² at 250 °C. This behavior is typical of Schottky diodes and is attributed to thermally excited electrons gaining sufficient energy to surmount the barrier or engage in thermionic field emission processes [59, 61, 65, 72]. The increase in leakage is particularly pronounced beyond 200 °C, where lattice vibrations and interfacial trap activity likely contribute to additional leakage channels. Consequently, the $I_{on}/I_{off}$ ratio decreases with increasing temperature, dropping by nearly two orders of magnitude from $1 \times 10^7$ at 25 °C to below $5.3 \times 10^5$ at 250 °C. Despite this decline, the device maintains strong rectification performance across the full temperature range, which indicates the thermal resilience of the LPCVD-grown β-$Ga_2O_3$ stack and the plasma-free etch-defined mesa architecture. These results confirm that the device remains functional and rectifying at elevated temperatures, and the interface quality is preserved even in thermally stressed regimes.

Finally, to assess the reverse blocking capability of the fabricated β-$Ga_2O_3$ Schottky diodes, reverse breakdown measurements were performed under steady-state voltage sweep conditions, as shown in Fig. 8(a). The diodes exhibited breakdown voltages ranging from 73 V to 100 V, corresponding to parallel-plate electric field strengths between 1.66 MV/cm and 1.94 MV/cm. The parallel-plate field at the reverse breakdown condition was estimated using one-dimensional electrostatics, $E_{\text{field}} = \sqrt{\frac{qN_dV_{\text{BR}}}{\varepsilon}}$, where q is the charge of the electron, $N_d$ is the doping of the semiconductor, $V_{BR}$ is the breakdown voltage, and $\varepsilon$ is the permittivity of $Ga_2O_3$. It should be noted that the observed breakdown voltages are primarily influenced by the relatively high doping concentration ($2.1 \times 10^{17}$ cm$^{-3}$) in the drift layer, which limits the depletion width and hence the maximum sustainable field. Although the LPCVD-grown β-$Ga_2O_3$ layers exhibit good structural and morphological quality, substrate and interface-related effects cannot be entirely ruled out. The use of c-plane sapphire, while advantageous for its cost-effectiveness and electrical insulation, introduces lattice mismatch with β-$Ga_2O_3$. This mismatch may lead to interfacial strain and dislocations that, although not obvious in surface morphology or crystallinity measurements, can influence local electric field distribution under reverse bias. Furthermore, the heteroepitaxial interface and mesa geometry may also result in localized field crowding and early onset of breakdown.

To further investigate the reverse breakdown behavior and electric field distribution, two-dimensional Technology Computer-Aided Design (TCAD) simulations were conducted based on the device geometry. The device cross-section in the simulation uses an inclined mesa sidewall (~60°), consistent with Fig. 1(c). As shown in Fig. 8(b), the simulated reverse bias field profile at 100 V reveals a pronounced electric field concentration near the perimeter of the anode contact. This effect is further confirmed by the extracted 1D electric field profile along the AA′ cutline in Fig. 8(c), which shows the peak electric field localized near the anode edge. The simulations also indicate that the lateral depletion spreading remains well confined within the anode-to-mesa edge spacing, indicating that sidewall effects are not the dominant limitation in this device geometry. These results align with common field crowding phenomena observed in vertical Schottky structures and emphasize the importance of careful peripheral contact layout and junction termination. Further improvements in field uniformity and breakdown performance could be achieved by incorporating optimized edge termination by deep etching or advanced field management designs in future device generations. Nevertheless, the results validate the feasibility of achieving consistent reverse blocking performance in quasi-vertical β-$Ga_2O_3$ Schottky diodes fabricated entirely through LPCVD growth and *in situ* plasma-free etching. This integration pathway offers a scalable, plasma-damage-minimized approach for developing future high-voltage β-$Ga_2O_3$ devices, particularly when paired with lower doping, thicker drift layers, engineered buffer architectures, and field management techniques, and represents an important first step in establishing a unified LPCVD growth-etch platform whose significance extends beyond the present device geometry to field-sensitive architectures, where the preservation of etched sidewall quality directly plays a critical role in controlling electric-field distribution, leakage, and breakdown.

## IV. Conclusions

In summary, this work demonstrates the first realization of quasi-vertical β-$Ga_2O_3$ Schottky barrier diodes fabricated entirely using a plasma-free LPCVD-based process, from epitaxial growth to *in situ* etching for mesa isolation, on insulating sapphire substrates. Leveraging *in situ* Ga-assisted LPCVD etching technique, we achieved plasma damage-free mesa isolation with etch depths of 3.6 μm and ~60° inclined sidewalls. The resulting SBDs showed excellent forward characteristics with an ideality factor of 1.29, a Schottky barrier height of 0.83 eV, and a low specific on-resistance of 8.6 mΩ·cm$^2$. Temperature-dependent I-V-T analysis revealed thermionic emission-dominated transport with strong thermal stability and a modest increase in Schottky barrier height and ideality factor, likely attributed to interface inhomogeneities and barrier non-uniformity. Devices exhibited low and stable reverse leakage characteristics and achieved breakdown fields up to 1.94 MV/cm. Although the breakdown voltage was limited by the relatively high drift layer doping concentration, the demonstrated reverse blocking capability, along with robust forward conduction and thermal performance, highlight the promise of LPCVD-grown and etched β-$Ga_2O_3$ devices. The integration of epitaxy and etching within a single LPCVD platform offers a scalable and low-damage fabrication route for next-generation UWBG power electronics. These results also provide a critical step toward enabling cost-effective, high-voltage, and thermally resilient β-$Ga_2O_3$-based device architectures on foreign substrates.

**Acknowledgement**

The authors acknowledge funding support from the National Science Foundation (NSF) under Award Nos. 2501623 and 2532898, the Ralph E. Powe Junior Faculty Enhancement Award (Fiscal Year 2025), sponsored by the Oak Ridge Associated Universities, the Draper Scholar Program and the Researchers and Scholars Investment Fund from the UMass Lowell Office of Research Development.

**Data Availability**

The data that support the findings of this study are available from the corresponding author upon reasonable request.

**Conflict of Interest**

The authors have no conflicts to disclose.

**Figure 1**

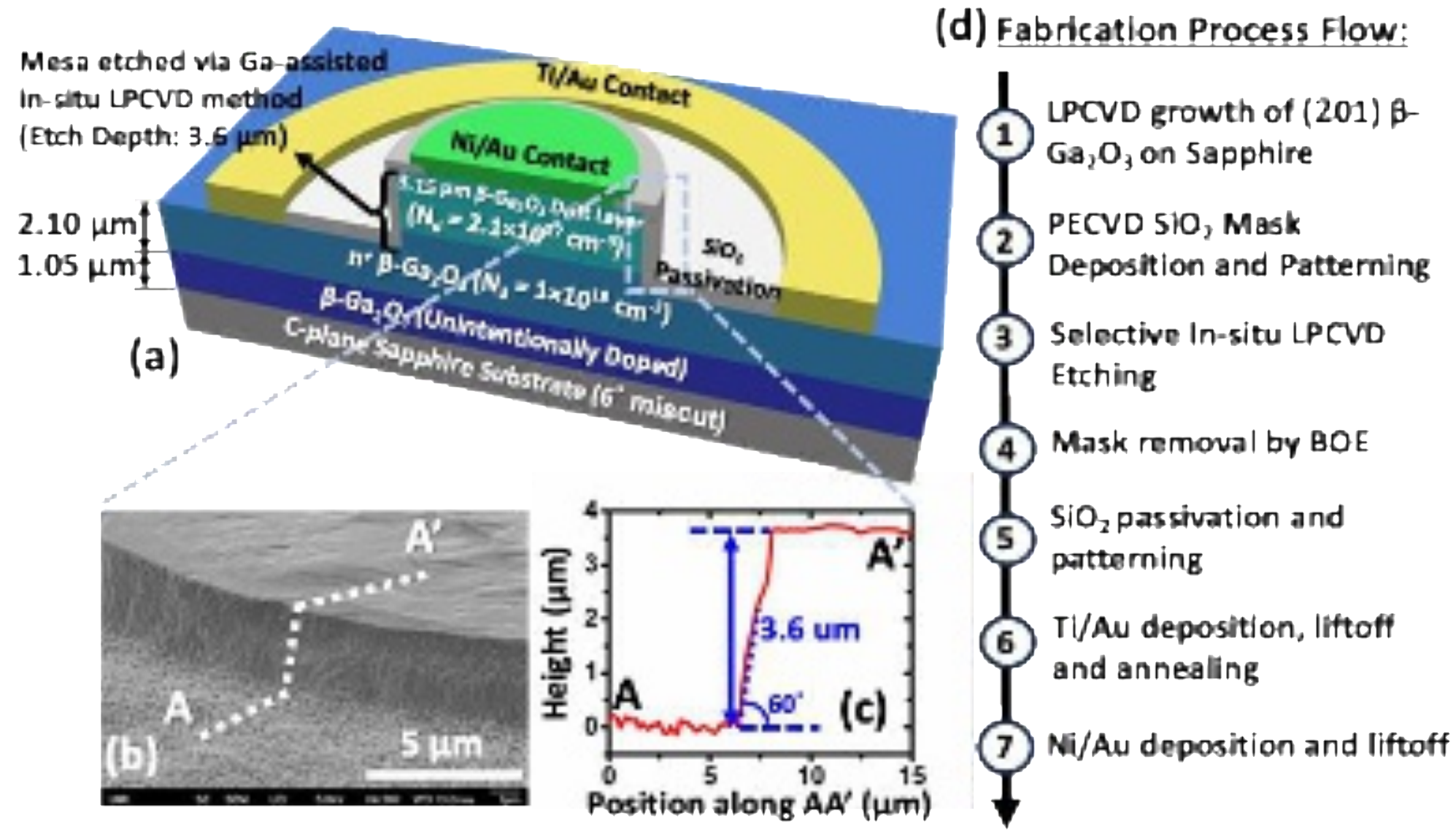

**Figure 1.** (a) Schematic cross-section of the quasi-vertical Schottky barrier diode (SBD) fabricated using LPCVD grown ($\bar{2}01$) β-$Ga_2O_3$ film on a 6° miscut c-plane sapphire substrate. The device features a three-layer epitaxial stack: a 1.05 µm unintentionally doped (UID) buffer layer, a 2.10 µm thick $n^+$ β-$Ga_2O_3$ contact layer ($N_d = 1 \times 10^{19}$ cm$^{-3}$), and a 3.15 µm β-$Ga_2O_3$ drift layer ($N_d = 2.1 \times 10^{17}$ cm$^{-3}$). Plasma-free LPCVD etching was used to etch 3.6 µm into the β-$Ga_2O_3$ drift layer for mesa isolation and to expose the $n^+$ layer for anode contact formation. (b) Tilted (60° tilt) Cross-sectional FESEM image showing the etch sidewall profile with an etch depth of 3.6 µm. (c) AFM line scan along the AA' cutline over etched mesa region confirming the ~60° sidewall inclination and an etch depth of 3.6 µm (d) Step by-step fabrication process flow including LPCVD growth, *in situ* etching, and metallization steps for device realization.

**Figure 2**

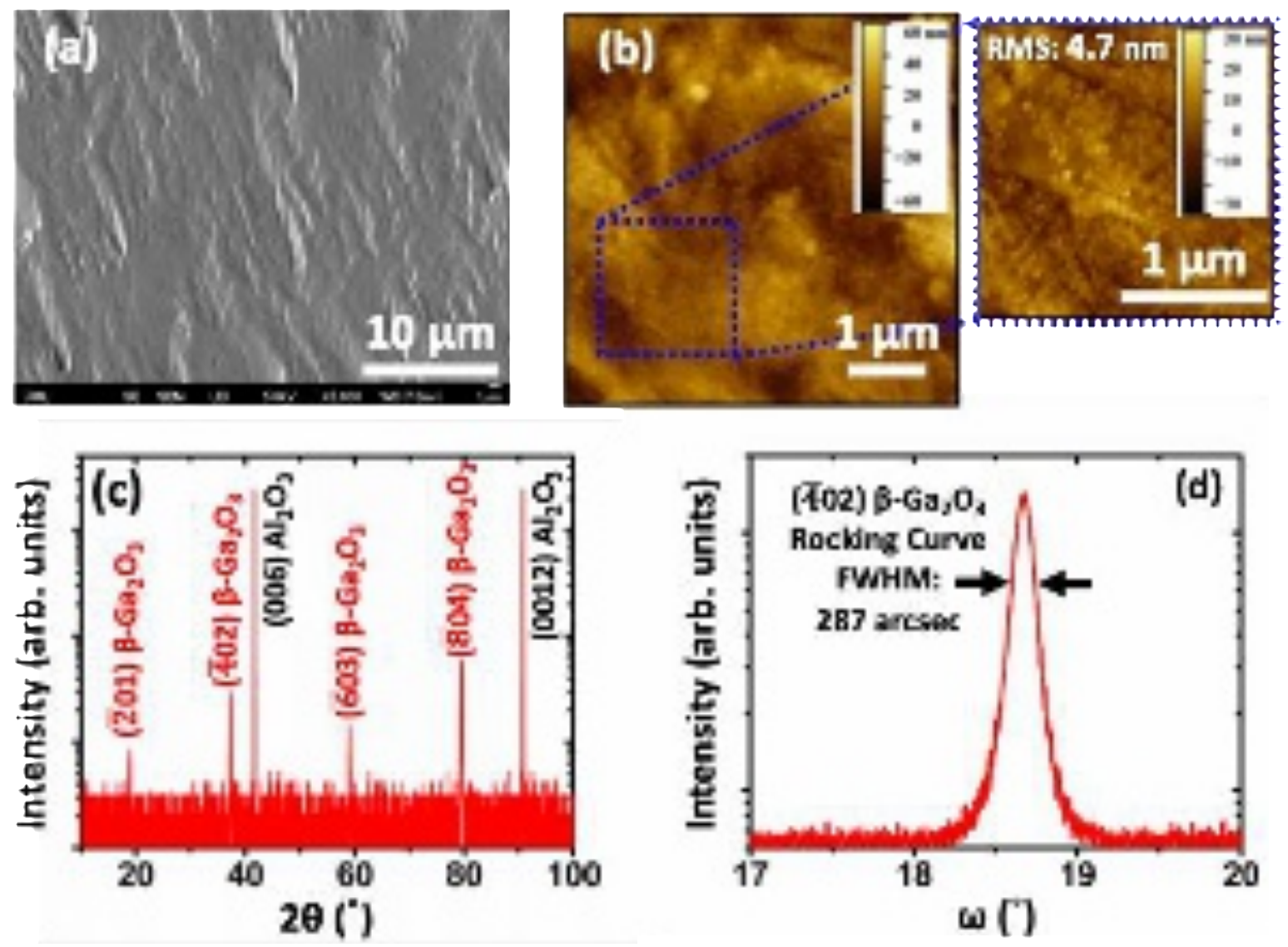

**Figure 2.** Structural and surface morphological characterization of the LPCVD-grown heteroepitaxial ($\bar{2}$01) β-$Ga_2O_3$ epitaxial stack used in device fabrication. (a) Top-view FESEM image showing clear step-flow surface morphology, indicative of high crystalline quality and epitaxial layer-by-layer growth. (b) AFM scan of the film surface reveals smooth morphology with a root mean square (RMS) roughness of 4.7 nm. (c) XRD ω-2θ scan showing strong diffraction peaks corresponding to the ($\bar{h}$0l) family of ($\bar{2}$01) β-$Ga_2O_3$, along with sapphire substrate peaks, confirming phase purity and preferred orientation. (d) XRD ω-rocking curve of the ($\bar{4}$02) reflection exhibiting a narrow full width at half maximum (FWHM) of 287 arcsec, indicating the high crystalline quality of the β-$Ga_2O_3$ epitaxial stack.

**Figure 3**

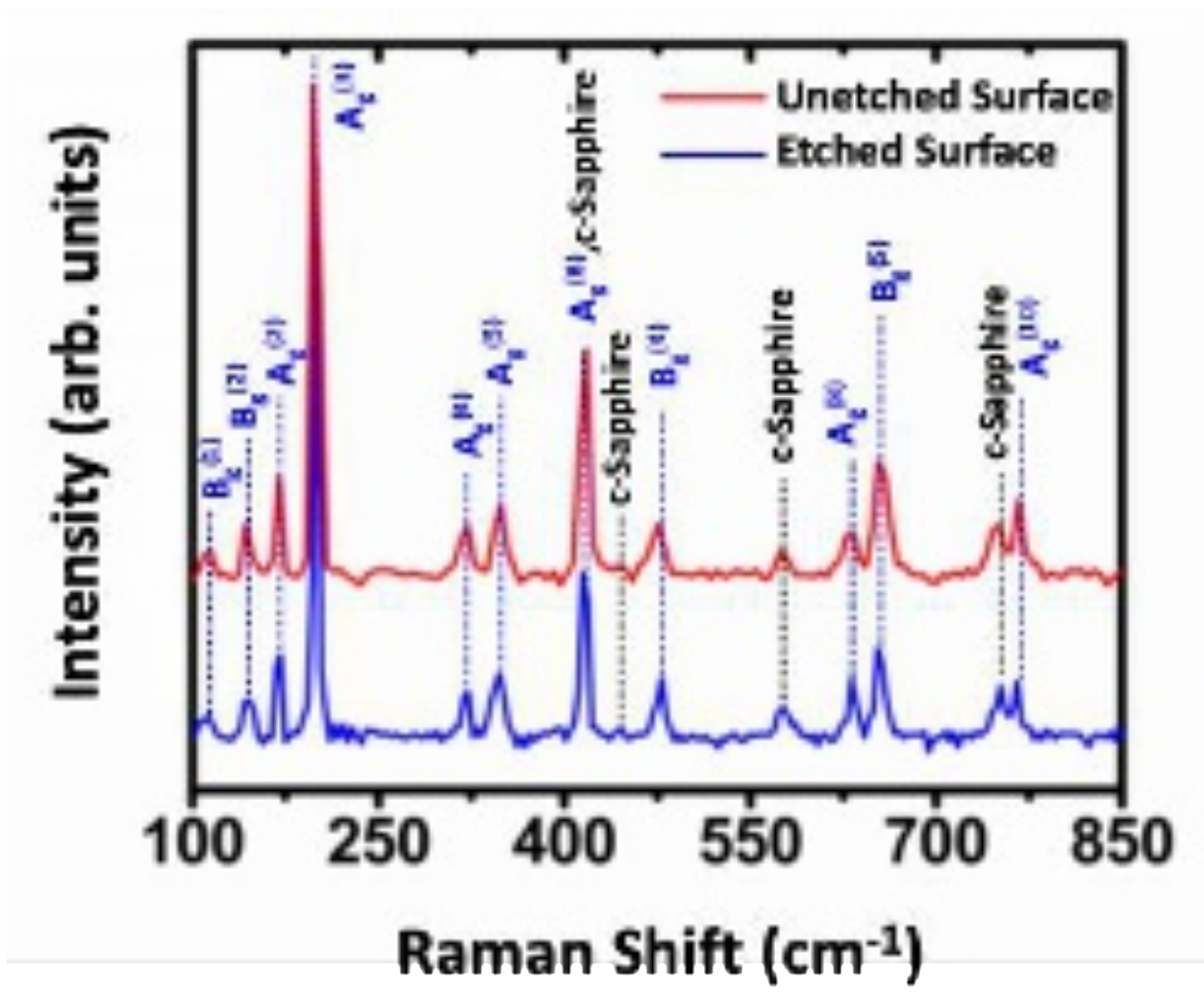

**Figure 3.** Room temperature Raman spectra of the LPCVD-grown heteroepitaxial ($\bar{2}01$) β-$Ga_2O_3$ surface before (red) and after (blue) Ga-assisted *in situ* LPCVD etching. Both spectra exhibit sharp and well-defined $A_g$ and $B_g$ phonon modes characteristic of monoclinic β-$Ga_2O_3$, confirming structural integrity and phase purity. Peaks from the c-plane sapphire substrate are also visible in both spectra.

**Figure 4**

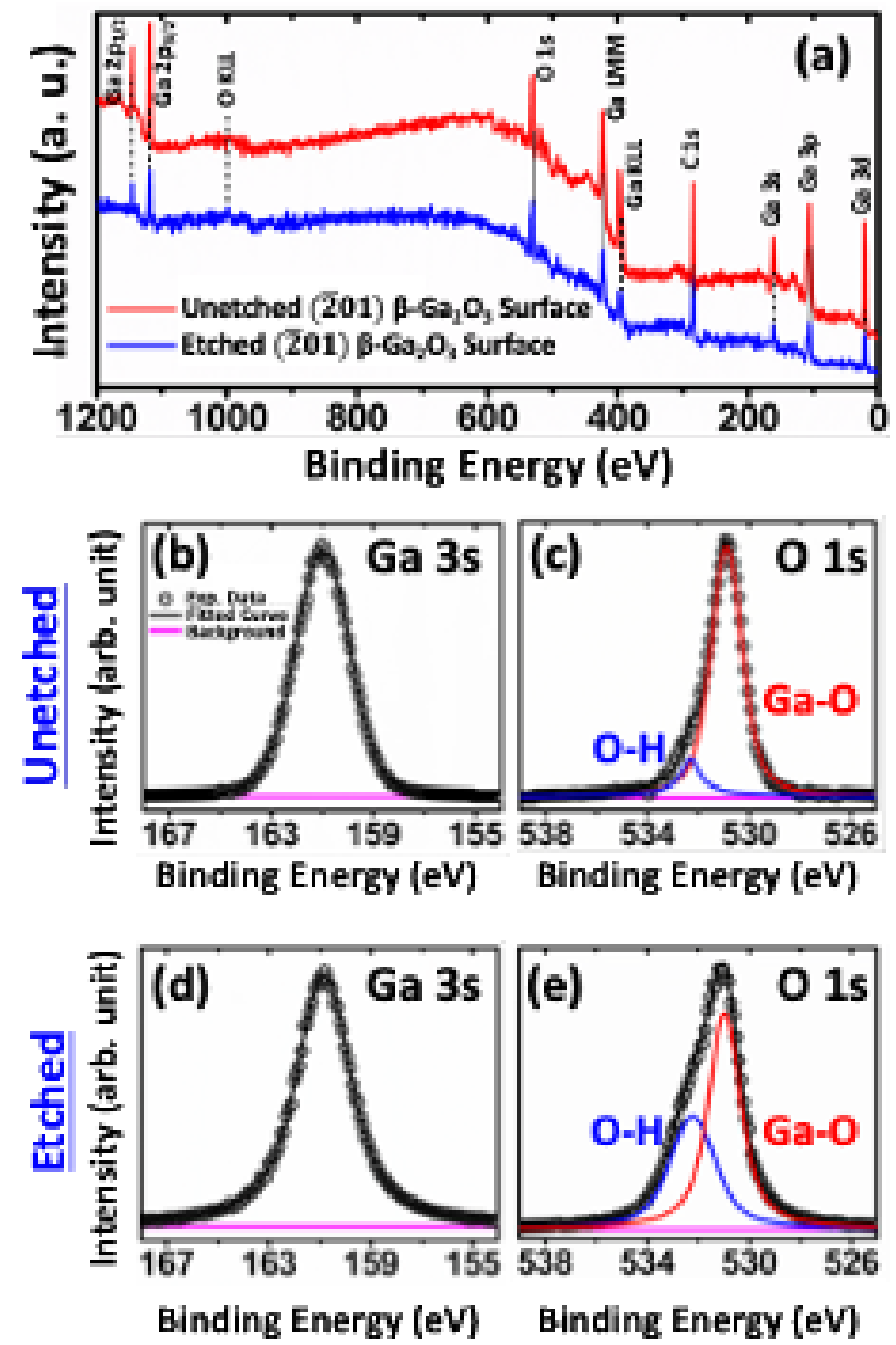

**Figure 4.** X-ray photoelectron spectroscopy analysis of LPCVD-grown $\beta$-$Ga_2O_3$ surfaces before and after Ga-assisted *in situ* etching. (a) Wide-scan XPS survey spectra showing Ga and O signals, with no detectable contamination from extrinsic elements. (b, c) High-resolution Ga 3s and O 1s spectra of the unetched surface. (d, e) Corresponding spectra for the etched surface. The O 1s peaks are deconvoluted into lattice oxygen (Ga-O) and surface hydroxyl (O-H) components. The spectra confirm chemical purity and near-stoichiometric composition for both surfaces, with no significant changes in bonding states after etching, indicating that the LPCVD etch process preserves surface chemistry and structural integrity.

**Figure 5**

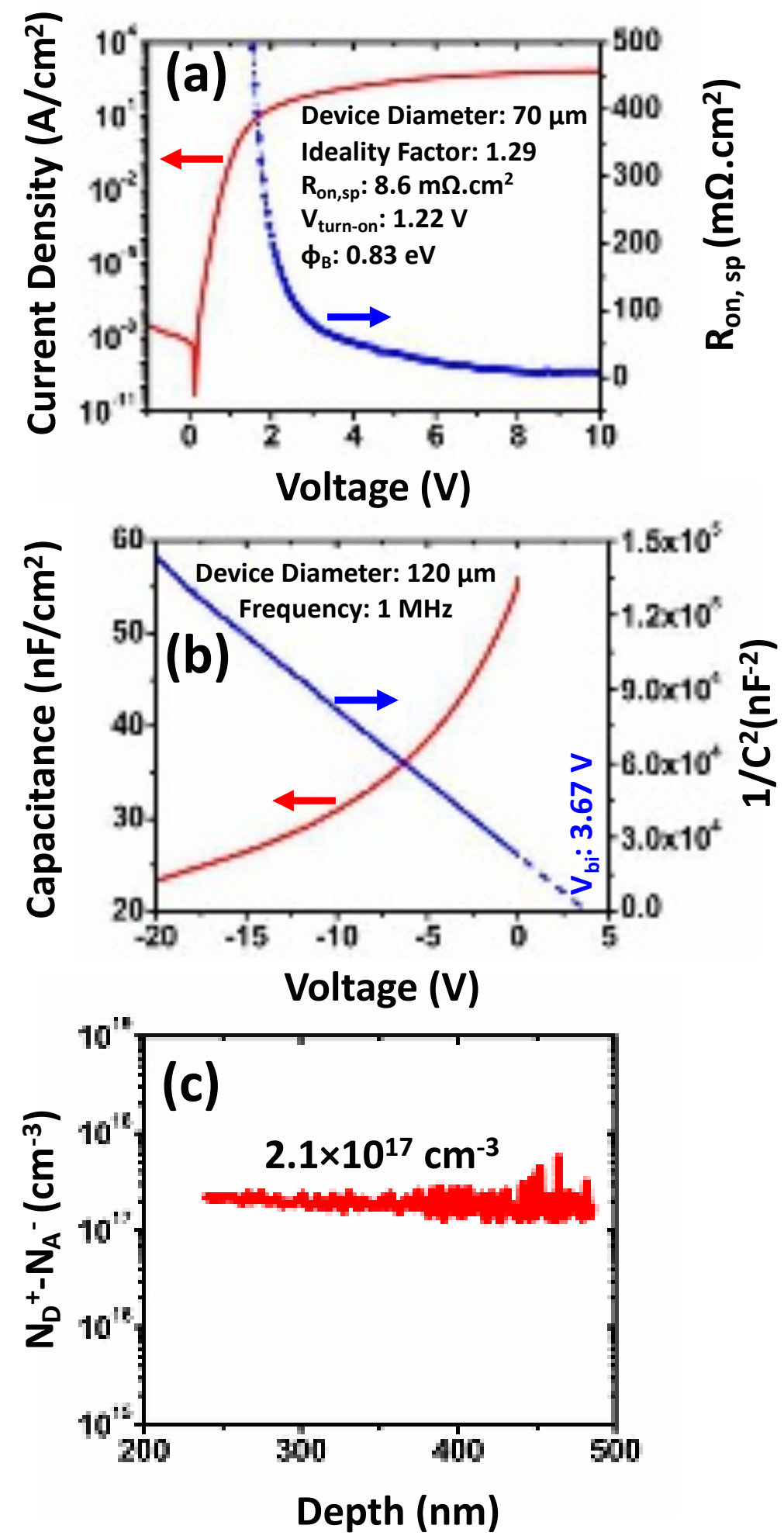

**Figure 5.** (a) Forward I-V characteristics and extracted specific on-resistance ($R_{on,sp}$) of the quasi-vertical $\beta$-$Ga_2O_3$ Schottky diode with a 70 µm diameter. The device exhibits an ideality factor of 1.29, a turn-on voltage ($V_{turn\text{-}on}$) of 1.22 V, a Schottky barrier height ($\Phi_B$) of 0.83 eV, and a minimum differential $R_{on,sp}$ of 8.6 mΩ·cm². (b) C-V and $1/C^2$-V characteristics measured at 1 MHz for a 120 µm diameter device. (c) Net carrier density profile ($Nd^+$-$Na^-$) showing uniform doping of $2.1 \times 10^{17}$ cm$^{-3}$ in the $\beta$-$Ga_2O_3$ drift layer, extracted from the C-V curve.

**Figure 6**

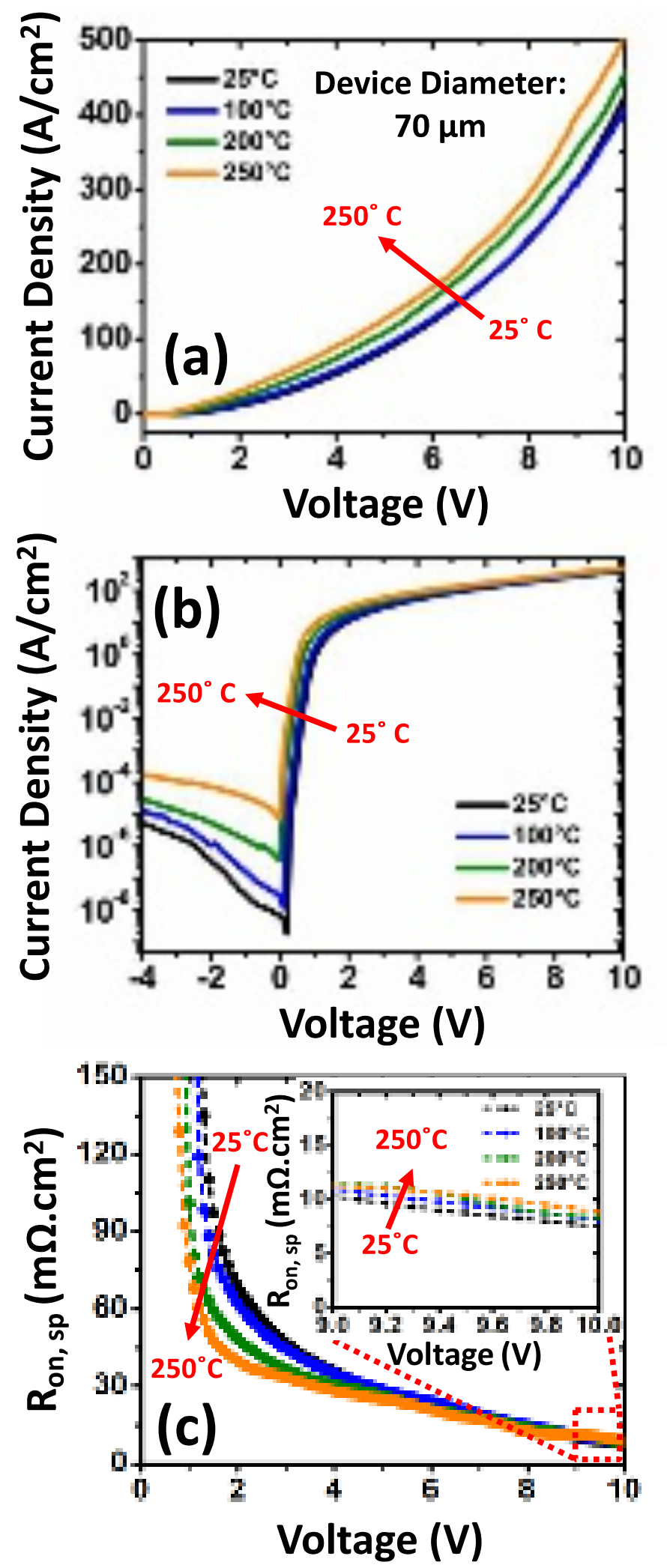

**Figure 6.** Temperature-dependent current-voltage (J-V-T) characteristics of a 70 µm diameter quasi-vertical β-$Ga_2O_3$ SBD measured at 25°C, 100°C, 200°C, and 250°C. (a) Forward J–V characteristics showing a monotonic increase in on-current density with temperature. (b) Semi-log J-V plot showing higher leakage current at elevated temperatures. (c) Extracted specific on-resistance ($R_{on,sp}$) versus forward bias voltage at different temperature.

**Figure 7**

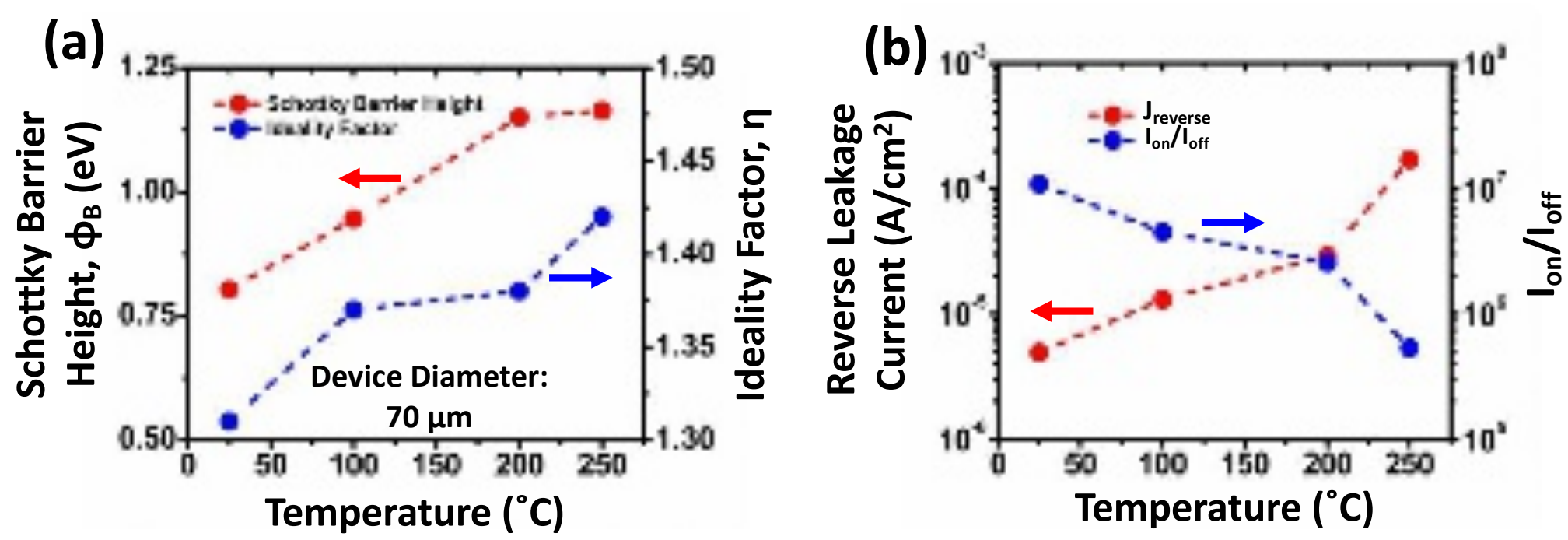

**Figure 7.** Temperature-dependent Schottky diode performance metrics. (a) Extracted Schottky barrier height ($\Phi_B$) and ideality factor (η) as a function of temperature, showing an increase in $\Phi_B$ and η with rising temperature. (b) Reverse leakage current ($J_{reverse}$) and on/off current ratio ($I_{on}/I_{off}$) versus temperature, illustrating the thermally activated increase in leakage and corresponding reduction in rectification ratio at elevated temperatures.

**Figure 8**

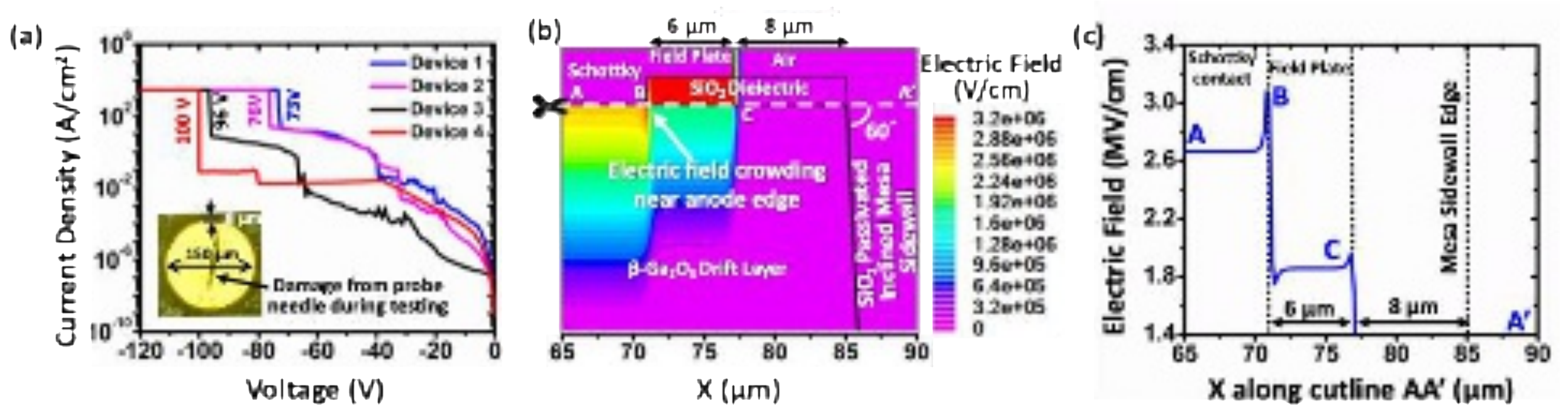

**Figure 8.** (a) Reverse J-V characteristics of four quasi-vertical β-$Ga_2O_3$ SBDs, showing breakdown voltages ranging from 73 V to 100 V, corresponding to parallel-plate electric field strengths between 1.66 MV/cm and 1.94 MV/cm. The inset shows an optical micrograph of a representative 150 µm-diameter device with an anode-to-mesa edge spacing of 8 µm. (b) TCAD-simulated 2D electric field distribution (contour plot) under reverse bias conditions (at 100V), revealing electric field crowding near the anode perimeter. The device cross-section in the simulation incorporates an inclined mesa sidewall (~60°), consistent with the measured device geometry. (c) 1D electric field profile extracted along the cutline AA′, further confirming peak field intensification near the anode edges.